\documentclass[fleqn,usenatbib]{mnras}

\usepackage{newtxtext,newtxmath}

\usepackage[T1]{fontenc}

\usepackage{graphicx}
\usepackage{amsmath}

\title[SMACS0723 star-formation histories at 5 < z < 9]{\vspace{-0.2cm}A first look at the SMACS0723 JWST ERO: spectroscopic redshifts, stellar masses and star-formation histories\vspace{-0.2cm}}

\author[A. C. Carnall et al.]
{A. C. Carnall$^{1}$\thanks{E-mail: adamc@roe.ac.uk},
R. Begley$^{1}$,
D. J. McLeod$^{1}$,
M. L. Hamadouche$^{1}$,
C. T. Donnan$^{1}$,
R. J. McLure$^{1}$,
\newauthor
J. S. Dunlop$^{1}$,
B. Milvang-Jensen$^{2, 3}$,
C. L. Bondestam$^{1}$,
F. Cullen$^{1}$,
S. M. Jewell$^{1}$,
C. L. Pollock$^{1}$
\\
$^{1}$ SUPA\thanks{Scottish Universities Physics Alliance}, Institute for Astronomy, University of Edinburgh, Royal Observatory, Edinburgh EH9 3HJ, UK\\
$^{2}$ Cosmic Dawn Center (DAWN)\\
$^{3}$ Niels Bohr Institute, University of Copenhagen, Jagtvej 128, 2200 Copenhagen, Denmark
\vspace{-0.5cm}}
\date{Accepted XXX. Received YYY; in original form ZZZ\vspace{-0.2cm}}
\pubyear{2022}

\begin{document}
\label{firstpage}
\pagerange{\pageref{firstpage}--\pageref{lastpage}}
\maketitle

\begin{abstract}
\noindent We present a first-look analysis of the JWST ERO data in the SMACS J0723.3-7327 cluster field. We begin by reporting 10 new spectroscopic redshifts from $\lambda_\mathrm{obs}=1.8-5.2\mu$m NIRSpec medium-resolution ($R=\lambda/\Delta\lambda = 1000$) data. These are determined via multiple high-SNR emission line detections, with 5 objects at $1 < z < 3$ displaying multiple rest-frame near-infrared Hydrogen Paschen lines, and 5 objects at $5 < z < 9$ displaying rest-frame optical Oxygen and Hydrogen Balmer lines. For the 5 higher-redshift galaxies we extract fluxes in 6 NIRCam bands spanning $\lambda_\mathrm{obs}=0.8-5\mu$m and perform spectral energy distribution fitting, in combination with existing HST photometry. The $7 < z < 9$ objects exhibit a U-shaped pattern across the F277W, F356W and F444W bands, indicating a Balmer break seen in emission (Balmer jump) and high-equivalent-width [O\,\textsc{iii}] emission. This indicates an extremely young stellar population, with the bulk of the current mass having formed within the past 10 Myr. We report robust stellar masses and mean stellar ages from our spectral fitting, with the four $z > 6$ galaxies exhibiting low stellar masses from log$_{10}(M_*/$M$_\odot)=7.1-8.2$ and correspondingly young mean stellar ages of only a few Myr. This work highlights the critical importance of combining large upcoming NIRCam surveys with NIRSpec follow-up to measure the spectroscopic redshifts necessary to robustly constrain physical parameters.
\end{abstract}
\begin{keywords}
Galaxies: high-redshift -- Galaxies: star formation -- Galaxies: distances and redshifts \vspace{-0.6cm}
\end{keywords}

\section{Introduction}

For much of the past decade, a ($\simeq500$ Myr) gap has existed in our knowledge of cosmic history, between the cosmic microwave background at redshift $z\simeq1100$, and the earliest known galaxies at $z\simeq10$ (e.g., \citealt{Coe2013, McLure2013, McLeod2016, Oesch2016, Donnan2022}). This has been largely due to a lack of deep, high-resolution imaging and spectroscopic capability at $\lambda > 2\mu$m. These instrumental limitations have also significantly restricted our knowledge of galaxy evolution during the first two billion years prior to $z=3$, due to our inability to study the detailed rest-frame {\it optical} properties of galaxies at such redshifts.

To constrain the build-up of stellar mass in currently unseen galaxies at $z>10$, much attention has focused on attempting to measure the star-formation histories (SFHs) of $6 < z < 10$ galaxies. The most important spectral feature is the Balmer break at $\lambda_\mathrm{rest}\simeq4000$\AA, which becomes stronger as galaxy stellar populations age, placing a lower bound on the redshift at which significant star formation commenced. The only data available for this purpose have been relatively shallow, low spatial resolution, very broad-band data from the \textit{Spitzer} IRAC 3.6$\mu$m and 4.5$\mu$m channels. The IRAC signature of a Balmer break can however be degenerate with strong [O\,\textsc{iii}]+H$\beta$ emission, especially when relying on uncertain photometric redshifts (e.g., \citealt{Oesch2015, Roberts-Borsani2016, Roberts-Borsani2020}).

\begin{figure*}
	\includegraphics[width=0.95\textwidth]{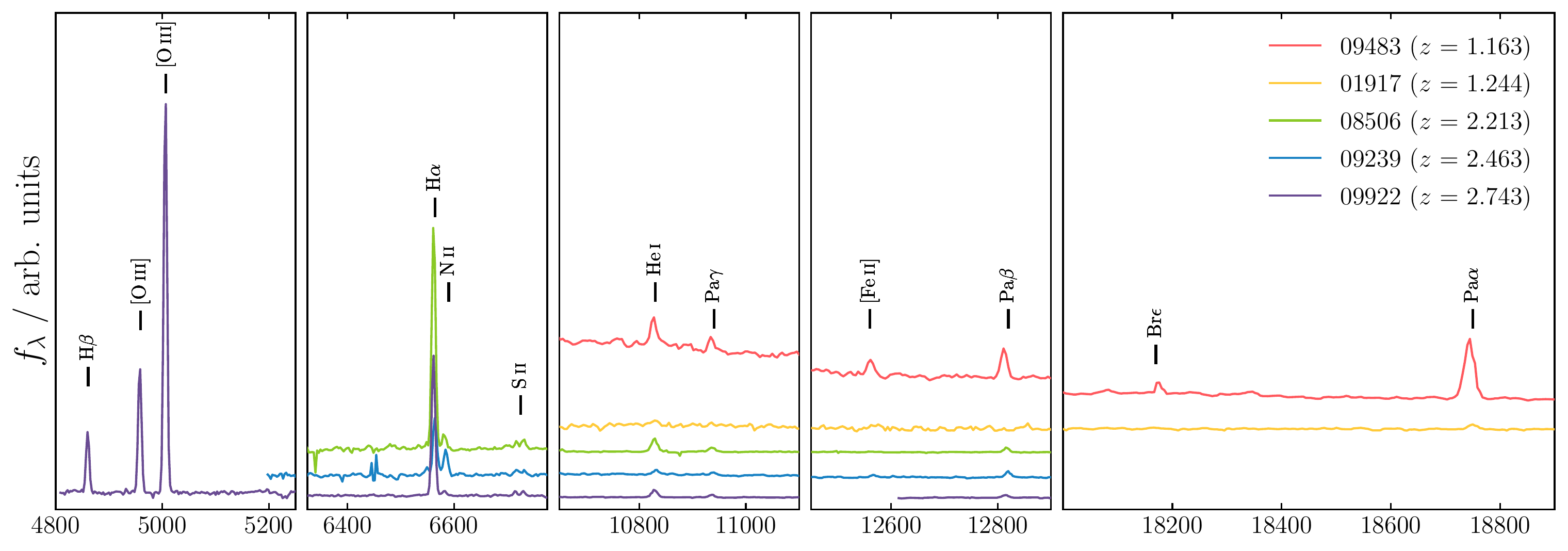}
	\includegraphics[width=0.95\textwidth]{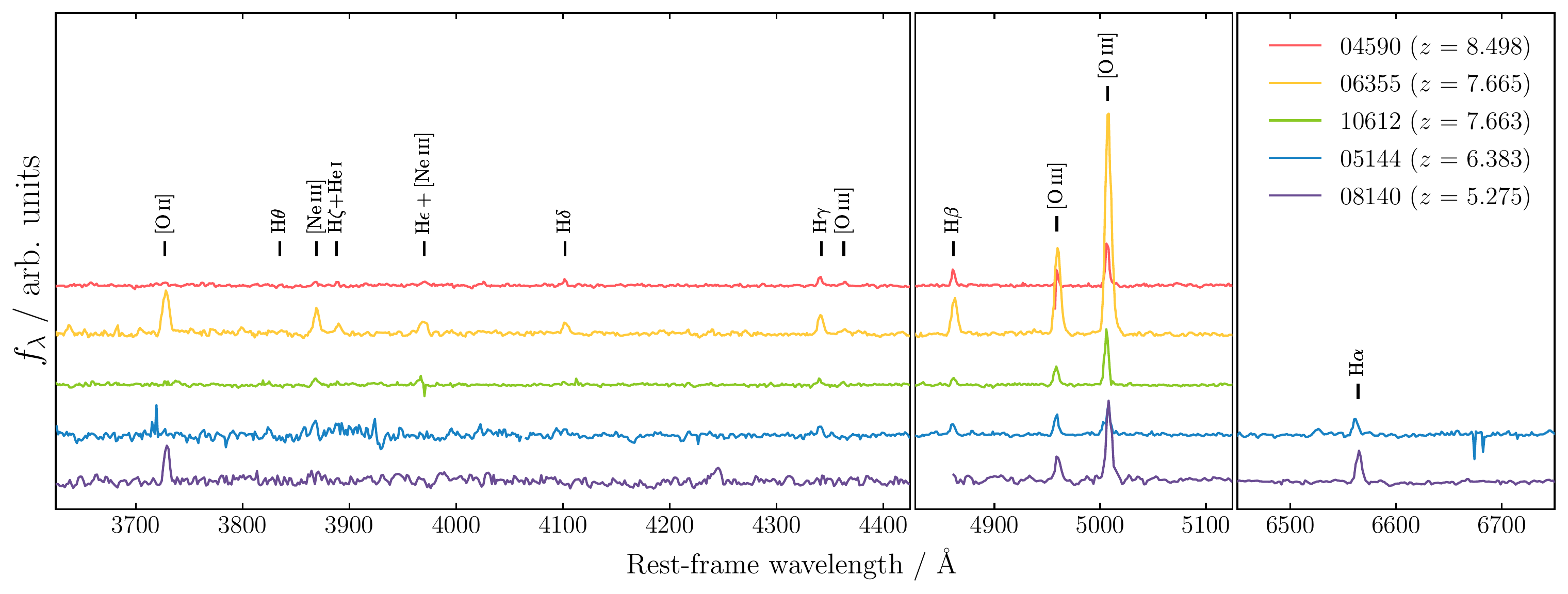}
    \caption{The 10 JWST NIRSpec spectra from which we were able to measure secure spectroscopic redshifts. Wavelength ranges containing key spectral features are excerpted from the full dataset ($\lambda_\mathrm{obs} \simeq1.8-5.2\mu$m). The top panels show objects at $1 < z < 3$, with redshifts determined primarily from Hydrogen Balmer, Paschen and Brackett lines. The lower panels show objects at $5 < z < 9$, with redshifts measured primarily from Oxygen and Hydrogen Balmer lines. The spectra have been flux normalised and vertical shifts applied for visualisation purposes. Missing sections are caused by gaps between the NIRSpec detectors.}
    \label{fig:spectra}
    \vspace{-0.4cm}
\end{figure*}

Recently, several authors have reported evidence for significant Balmer breaks in the spectra of galaxies at $z\simeq8-9$ (e.g., \citealt{Hashimoto2018, Strait2020, Strait2021, Laporte2021}). These results suggest stellar populations with ages of several hundred Myr already in place when the Universe was only $\simeq600$ Myr old, in some cases implying that significant star formation was underway as early as $\simeq100$ Myr after the Big Bang ($z\simeq30$). However, constraining galaxy SFHs from photometric data is challenging due to the age-metallicity-dust degeneracy in galaxy spectral shapes (e.g., \citealt{Conroy2013}), as well as the ill-conditioned nature of the galaxy spectral fitting problem, which results in strong prior-dependence (e.g., \citealt{Ocvirk2006, Carnall2019a, Leja2019}). The above considerations led \cite{Tacchella2022} to conclude that stellar ages for $z\simeq10$ galaxies derived from current data are still highly uncertain.

In \cite{Whitler2022}, the authors find a range of different SFHs for $z=6.6-6.9$ galaxies. They suggest that the most luminous objects at this epoch are a mixture of the most massive galaxies, with ages of up to a few hundred Myr, and galaxies that have undergone a very recent, rapid increase in star formation during the preceding $\lesssim10$\,Myr. However, at these redshifts the wavelength range of interest is only sampled by one, very broad \textit{Spitzer} IRAC band. There is also no coverage of the wavelength range $\lambda_\mathrm{obs} =2.3-3.1\mu$m. To make further progress, separating out the Balmer break from extreme-equivalent-width line emission in the rest-frame optical is critical.

By providing ultra-deep, high spatial and spectral resolution imaging and spectroscopy as far into the infrared as $\lambda_\mathrm{obs} = 30\mu$m, including particularly wide-ranging capabilities at $1-5\mu$m, the \textit{James Webb Space Telescope} (JWST) is set to revolutionise our understanding of galaxy formation during the first few billion years of cosmic history. This will allow us not only to reliably detect, and confirm redshifts for, large samples of $z>10$ galaxies, but also to gain a detailed physical understanding of galaxies at $3 < z < 10$ (e.g., \citealt{Chevallard2019, Kemp2019, Roberts-Borsani2021}).

\begin{table}
  \caption{Redshifts for the 5 galaxies shown in the top panels of Fig. \ref{fig:spectra}.}
\label{table:redshifts_loz}
\begingroup
\setlength{\tabcolsep}{12pt} 
\begin{center}
\begin{tabular}{lccc}
\hline
ID & Redshift & RA & DEC \\
\hline
1917 & 1.244 & 110.87105 & $-$73.46559 \\
8506 & 2.213 & 110.91640 & $-$73.45864 \\
9239 & 2.463 & 110.76578 & $-$73.45161 \\
9483 & 1.163 & 110.79735 & $-$73.44899 \\
9922 & 2.743 & 110.85947 & $-$73.44409 \\
\hline
\vspace{-1cm}
\end{tabular}
\end{center}
\endgroup
\end{table}

In this paper, we focus on the first publicly released data from JWST, the Early Release Observations (ERO; \citealt{Pontoppidan2022}) covering the SMACS J0723.3–7327 galaxy cluster (hereafter SMACS0723). We aim to demonstrate the improvement in galaxy physical parameter constraints that can be achieved at $5 < z < 9$ by combining spectroscopic redshifts from NIRSpec with deeper, redder and narrower-band photometry from NIRCam.

We begin by reporting 10 spectroscopic redshifts from a total of 35 objects that were observed with the NIRSpec microshutter array (MSA) ($\lambda_\mathrm{obs}=1.8-5.2\mu$m, at spectral resolution $R=1000$). Five of these objects are at $1 < z < 3$, with redshifts measured principally via Hydrogen Paschen lines. The other 5 span $5 < z < 9$, with their redshifts determined from strong rest-frame optical Oxygen and Hydrogen Balmer lines. For the 5 high-redshift galaxies, we measure fluxes in the 6 NIRCam bands included in the ERO, spanning $\lambda_\mathrm{obs}=0.8-5\mu$m. We perform spectral fitting with \textsc{Bagpipes} \citep{Carnall2018}, employing our new spectroscopic redshifts and JWST photometric data, in combination with existing \textit{Hubble Space Telescope} (HST) photometry. We measure stellar masses, with these objects being some of the first for which robust masses are available at these redshifts. We also discuss the SFHs of these objects, with the aim of constraining the redshifts at which they began forming stars.

The structure of this paper is as follows. In Section \ref{data} we introduce the NIRCam and NIRSpec data for SMACS0723. In Section \ref{redshifting}, we describe the redshift measurements from the NIRSpec data. In Section \ref{sed_fitting} we present our spectral energy distribution (SED) fitting methodology and results. We present our conclusions in Section \ref{conclusion}. All magnitudes are quoted in the AB system. For cosmological calculations, we adopt $\Omega_M = 0.3$, $\Omega_\Lambda = 0.7$ and $H_0$ = 70 $\mathrm{km\ s^{-1}\ Mpc^{-1}}$. We assume a \cite{Kroupa2001} initial mass function, and assume the Solar abundances of \cite{Asplund2009}, such that $\mathrm{Z_\odot} = 0.0142$.\vspace{-0.5cm}

\section{Data}\label{data}

\subsection{NIRCam imaging}\label{data_nircam}

All JWST observations used in this work were taken as part of the SMACS0723 ERO (Programme ID 2736). We utilise deep NIRCam imaging in the F090W, F150W, F200W, F277W, F356W and F444W filters, providing coverage of the $\lambda_\mathrm{obs} = 0.8-5\mu$m wavelength range.
We reduce the raw level-1 data products using PENCIL (PRIMER Enhanced NIRCam Image Processing Library), a custom version of the JWST pipeline (version 1.6.2), using the latest available calibration files (CRDS\_CTX =  jwst\_0984.pmap). We align and stack the reduced images using a combination of {\sc scamp} \citep{Bertin2006} and {\sc swarp} \citep{Bertin2010}, producing final deep images aligned to GAIA EDR3 \citep{Gaia2021} with a pixel scale of 0.031$^{\prime\prime}$. We also make use of HST ACS F606W and F814 data, starting with the mosaics made available by the Reionization Lensing Cluster Survey (RELICS; \citealt{Coe2019}) team. We PSF-homogenise each band to the F444W filter using an empirical PSF, derived by stacking bright stars. We then extract photometric fluxes in $0.5^{\prime\prime}-$diameter apertures. We apply a calibration correction to our NIRcam fluxes, derived as described in Appendix C of \cite{Donnan2022}. To measure robust photometric uncertainties, we measure the aperture-to-aperture rms of the nearest $\sim 200$ blank sky apertures, after masking out neighbouring sources \citep{McLeod2016}.\vspace{-0.5cm}

\subsection{NIRSpec spectroscopy}\label{data_nirspec}

Two NIRSpec MSA pointings were conducted, s007 and s008, with galaxies selected as described in \cite{Pontoppidan2022}. For each pointing, two grism/filter combinations were used: G235M/F170LP and G395M/F290LP, providing coverage over the wavelength range $\lambda_\mathrm{obs} \simeq1.8-5.2\mu$m at spectral resolution $R=\lambda/\Delta\lambda\simeq1000$. The 10 objects discussed in this work all received the full integration time of 8754 seconds in both pointings and with both grism/filter combinations. We have used the original level-3 data products made available on 12/07/2022, which were processed with version 1.5.3 of the JWST Science Calibration Pipeline. The calibration reference data used was jwst\_0916.pmap. Coordinates for each object were obtained by cross-matching IDs with the Astronomer's Proposal Tool (APT) input catalogue and refined using the NIRCam imaging.\vspace{-0.5cm}

\begin{table*}
  \caption{Redshifts, stellar masses and mean stellar ages for the 5 high-redshift galaxies for which spectroscopic redshifts could be measured. The lensing factors were taken from the maps published by the RELICS team using the \textsc{Glafic} tool \citep{Oguri2010}. The SEDs and SFHs for these galaxies are shown in Fig. \ref{fig:sfhs}.\vspace{-0.2cm}}
\label{table:redshifts_hiz}
\begingroup
\setlength{\tabcolsep}{12pt} 
\renewcommand{\arraystretch}{1.1} 
\begin{tabular}{lcccccc}
\hline
ID & Redshift & RA & DEC & log$_{10}(M_*/$M$_\odot)$ & Mean stellar age / Myr & Lensing factor\\
\hline
4590 & 8.498 & 110.85933 & $-$73.44916 & $7.10^{+0.14}_{-0.12}$ & $1.3^{+1.3}_{-0.3}$& 10.09 \\
5144 & 6.383 & 110.83972 & $-$73.44536 & $7.39^{+0.03}_{-0.04}$ & $1.2^{+0.5}_{-0.2}$ & 2.89 \\
6355 & 7.665 & 110.84452 & $-$73.43508 & $8.23^{+0.08}_{-0.09}$ & $1.3^{+0.4}_{-0.3}$ & 2.69 \\
8140 & 5.275 & 110.78804 & -73.46179 & $8.72^{+0.20}_{-0.24}$ & $16^{+19}_{-9}$ & 1.67 \\
10612 & 7.663 & 110.83395 & $-$73.43454 & $7.72^{+0.06}_{-0.05}$ & $1.2^{+0.3}_{-0.2}$ & 1.58 \\
\hline
\vspace{-0.7cm}
\end{tabular}
\endgroup
\end{table*}

\section{Spectroscopic redshift determination}\label{redshifting}

The spectra described in Section \ref{data_nirspec} were redshifted by a combination of visual inspection and the Pandora.ez tool \citep{Garilli2010}. From the 35 objects for which data are available, secure redshifts could be obtained in 10 cases. In each case, a range of high-SNR emission line detections were observed, leading to precise and unambiguous spectroscopic redshifts. Key sections of the spectra are shown in Fig. \ref{fig:spectra}, with emission features labelled. Object IDs, coordinates and spectroscopic redshifts are presented in Tables \ref{table:redshifts_loz} and \ref{table:redshifts_hiz}.

The galaxies for which redshifts could be obtained fall into two categories. Firstly, the 5 objects shown in the top panel of Fig. \ref{fig:spectra} fall within the redshift range $1 < z < 3$, with their redshifts determined primarily from Hydrogen Balmer, Paschen and Brackett lines. These near-infrared Hydrogen lines hold much promise as star-formation-rate indicators, as they are far less affected by dust than H$\alpha$ (e.g., \citealt{Pasha2020}). The clear detection of these lines showcases the unique capabilities of JWST. Another highlight is the detection of He\,\textsc{i} 10830\AA\ (e.g., \citealt{Groh2007}) and [Fe\,\textsc{ii}] 12570\AA\ (e.g., \citealt{Izotov2009}), both of which are associated with massive stars.

The focus of this work however is on the second group, shown in the bottom panels of Fig. \ref{fig:spectra}, which comprises $5 < z < 9$ galaxies. For these objects, redshifts were measured primarily using a combination of rest-frame optical Hydrogen Balmer and Oxygen lines. Interestingly, two objects (6355 and 10612) display almost identical redshifts. These objects are however far from the cluster centre, and are not listed as multiple images in currently available lensing analyses of this field (\citealt{Mahler2022, Pascale2022, Caminha2022}). Several spectra show clear detections of the [O\,\textsc{iii}] 4363\AA\ auroral line, commonly used in ``direct" method metallicity measurements \citep{Kewley2019}.

\begin{figure*}
	\includegraphics[width=\textwidth]{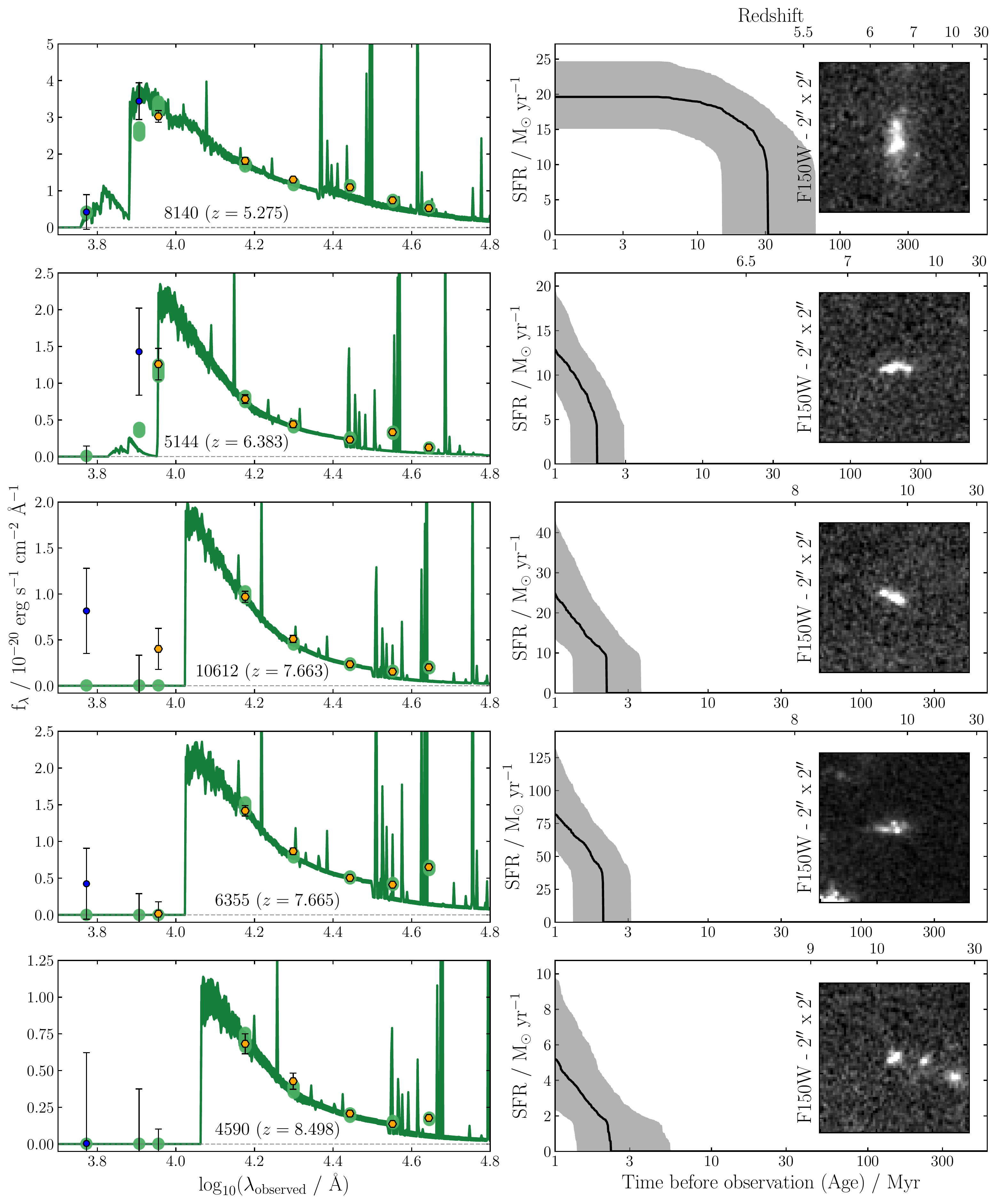}
    \caption{Spectral energy distributions, SFHs and F150W cutout images for the five SMACS0723 galaxies at $5 < z < 9$ with spectroscopic redshifts. In the left panels, blue circles indicate HST photometry, whereas golden hexagons indicate JWST NIRCam photometry. The three highest-redshift objects exhibit a characteristic U-shaped pattern in the F277, F356 and F444W bands indicative of a Balmer break seen in emission and high-equivalent-width [O\,\textsc{iii}]+H$\beta$ emission. This indicates a large increase in SFR within the last 10 Myr. Only the lowest-redshift object in the top panel exhibits a traditional Balmer break.}
    \label{fig:sfhs}
\end{figure*}

\section{Spectral Energy Distributions}\label{sed_fitting}

The HST+NIRCam SEDs for the five $5 < z < 9$ galaxies are shown in the left-hand panels of Fig. \ref{fig:sfhs}. At $z\simeq7-9$, the F277W and F356W filters ($4.4< \log_{10}(\lambda/$\AA$)<4.6)$ bracket the Balmer break, whereas [O\,\textsc{iii}] 5007\AA\ is in the F444W filter. For our three highest-redshift objects, a U-shaped pattern can be seen across these three filters. This indicates that the Balmer break is seen in emission, sometimes referred to as a Balmer jump. This is a signature of a galaxy dominated by a very young stellar population, with the additional flux below the break provided by nebular continuum emission and (potentially) by extremely massive stars \citep{Martins2020}. Evidence for a traditional Balmer break, and hence an older stellar population, is seen only for our lowest-redshift object, at $z=5.275$, with excess flux in the F277W filter. However, it should be noted that [O\,\textsc{iii}] 5007\AA\ falls into the edge of F277W at this redshift, in a region where the filter transmission is approximately a third of its maximum value.

To better understand the SFHs of these galaxies, their SEDs were fitted using \textsc{Bagpipes} \citep{Carnall2018}. Due to early highly uncertain flux calibration of the spectroscopic data, we use only the photometric data described in Section \ref{data_nircam}, whilst fixing redshifts to the spectroscopic values in Table \ref{table:redshifts_hiz}. We use the 2016 updated version of the \cite{Bruzual2003} stellar population models with the MILES stellar spectral library. We allow the logarithm of the stellar metallicity, $Z_*$, to vary with a uniform prior from $-2 < \log_{10}(Z_*/$Z$_\odot) < -0.3$. Nebular emission is included via the \textsc{Cloudy} code \citep{Ferland2017}, following the method in \cite{Carnall2018}. We allow the logarithm of the ionization parameter to vary over the range $-2 <$ log$_{10}(U) < -4$ with a uniform prior. We model dust attenuation with the \cite{Salim2018} model, using the same priors as \cite{Carnall2020}. We vary the ratio of attenuation between stars in stellar birth clouds (assuming 10 Myr lifetime) and the broader interstellar medium with a uniform prior from 1 to 3.

We explored a variety of different SFH models (e.g. exponentially rising, double power law) to try to understand the constraining power of these new data. In all cases, the four highest-redshift galaxies could not be well fitted except by models in which the bulk of the current stellar population formed within the preceding 10\,Myr. In particular, this is necessary to reproduce the Balmer jump between the F277W and F356W bands seen in the three highest-redshift spectra. In the end, we use a simple constant SFH model, which is adequate to explain the data. We vary the age using a logarithmic prior from 1 Myr to the age of the Universe.

The results of our SED fitting analysis are also shown in Fig. \ref{fig:sfhs}, with key parameters listed in Table \ref{table:redshifts_hiz}. We correct our SFHs and stellar masses using the lensing model released by the RELICS team computed with \textsc{Glafic} \citep{Oguri2010}. We find very young ages for the four highest-redshift objects, significantly below 10 Myr. This is perhaps unsurprising however, given their relatively low delensed stellar masses. The highest mass, SFR and most extreme Balmer jump all belong to object 6355, which shows clear structure in the F150W imaging, potentially indicative of an ongoing merger event.

Our three $z>7$ galaxies were also recently studied by \cite{Tacchella2022}, who recover median age estimates of $3-7$ Myr using non-parametric SFHs. It is well known that parametric SFH models, as employed in this work, typically produce relatively young ages (e.g. \citealt{Wuyts2011, Carnall2019a}), whereas non-parametric models typically produce older estimates (e.g. \citealt{Panter2007, Leja2019}). These slightly older ages are therefore expected given the methodological differences between our studies, and indicate a high probability that these galaxies are significantly younger than 10 Myr, in accord with recent \hbox{theoretical predictions \citep{Mason2022}.}
\section{Conclusion}\label{conclusion}

In this work we present a first-look analysis of the SMACS0723 JWST ERO data, focusing on galaxies with new spectroscopic redshifts from NIRSpec, in particular those at $5 < z < 9$. We report 10 new redshifts from the ERO NIRSpec data, which are shown in Fig. \ref{fig:spectra}. Half of these spectra are for comparatively low redshift ($1 < z < 3$) galaxies, for which NIRSpec detects a wealth of rest-frame near-infra-red emission lines, primarily from the Hydrogen Paschen series. The other 5 spectra are for $5 < z < 9$ galaxies, which display rest-frame optical Hydrogen Balmer and Oxygen lines.

We then fit SEDs generated from HST+NIRCam imaging data for the 5 high-redshift galaxies, focusing on determining their stellar masses and SFHs. For the four $z>6$ objects we see evidence for a Balmer break in emission (Balmer jump), associated with a very young stellar population, the bulk of which must have formed within the past 10 Myr. The three highest-redshift galaxies in particular show a U-shaped pattern in the F277W, F356W and F444W bands, due to the presence of the Balmer jump and high-equivalent-width [O\,\textsc{iii}]+H$\beta$ emission. These extremely young ages are consistent with the relatively low stellar masses we find for these galaxies, with all except the lowest-redshift ($z=5.275$) being comfortably below log$_{10}(M_*/$M$_\odot) = 9$ when corrected for lensing effects.

Larger-area JWST surveys, such as Cosmic Evolution Early Release Science (CEERS) and Public Release IMaging for Extragalactic Research (PRIMER), may well uncover more-mature and more-massive galaxies at $z>7$ that do contain stellar populations old enough to exhibit clear Balmer breaks in NIRCam imaging. However, this study highlights the key importance of coupling such imaging surveys with deep NIRSpec follow-up observations, in order to obtain the robust spectroscopic redshifts necessary to distinguish between Balmer breaks and high-equivalent-width [O\,\textsc{iii}]+H$\beta$ emission.\vspace{-0.5cm}

\section*{Acknowledgements}

ACC thanks the Leverhulme Trust for their support via a Leverhulme Early Career Fellowship. RB, DJM, MLH, CD, RJM, JSD and FC acknowledge the support of the Science and Technology Facilities Council. SJ and CP acknowledge the support of the School of Physics \& Astronomy, University of Edinburgh via Summer Studentship bursaries. The Cosmic Dawn Center is funded by the Danish National Research Foundation under grant no. 140.\vspace{-0.5cm}

\section*{Data Availability}

All JWST and HST data products are available via the Mikulski Archive for Space Telescopes (\url{https://mast.stsci.edu}). \vspace{-0.5cm}

\bibliographystyle{mnras}
\bibliography{carnall2022b}

\label{lastpage}
\end{document}